\documentclass[aps,prl,floatfix,amsmath,amssymb,nofootinbib,twocolumn]{revtex4-1}
\usepackage{graphicx,color,dcolumn,booktabs,bm,adjustbox,esint,upgreek}
\usepackage{longtable,comment,braket}
\usepackage{amsfonts,mathrsfs}
\usepackage{times}
\usepackage{overpic}
\usepackage{tipa}
\usepackage{indentfirst}
\usepackage{feynmf}   
\usepackage{slashed}  
\usepackage{cases}
\usepackage{epstopdf}
\usepackage{psfrag}
\usepackage{subfigure}
\usepackage{capt-of}
\usepackage{dcolumn,kantlipsum}
\usepackage{color}
\usepackage{multirow}
\usepackage{ulem}
\usepackage[colorlinks, citecolor=blue,anchorcolor=red,menucolor=red,linkcolor=blue,filecolor=red,runcolor=red,urlcolor=blue,frenchlinks=red]{hyperref}
\allowdisplaybreaks[4]

\begin{document}
\title{Deciphering the charged heavy quarkoniumlike states in chiral effective field theory}
\author{Bo Wang$^{1,2}$}\email{bo-wang@pku.edu.cn}
\author{Lu Meng$^{1,2}$}\email{lmeng@pku.edu.cn}
\author{Shi-Lin Zhu$^{1,2}$}\email{zhusl@pku.edu.cn}
\affiliation{
$^1$Center of High Energy Physics, Peking University, Beijing 100871, China\\
$^2$ School of Physics and State Key Laboratory of Nuclear Physics
and Technology, Peking University, Beijing 100871, China}

\begin{abstract}
We generalize the framework of chiral effective field theory to
study the interactions of the isovector $D^\ast\bar{D}^{(\ast)}$ and
$B^\ast\bar{B}^{(\ast)}$ systems up to the next-to-leading order, in
which the long-, mid-, and short-range force contributions as well
as the $S$-$D$ wave mixing are incorporated. Based on the
Lippmann-Schwinger equation, we fit the invariant mass distributions
of the elastic channels measured by the BESIII and Belle
Collaborations. Our results indicate that the four charged
charmoniumlike and bottomoniumlike states $Z_c(3900)$, $Z_c(4020)$
and $Z_b(10610)$, $Z_b(10650)$ can be well identified as the
$D\bar{D}^{\ast},D^\ast\bar{D}^{\ast}$ and
$B\bar{B}^{\ast},B^\ast\bar{B}^{\ast}$ molecular resonances. The
bound state explanations are vetoed in our framework. Our study
favors the $Z_c$ and $Z_b$ states are the twin partners under the
heavy quark symmetry.
\end{abstract}

\maketitle

Hadrons are usually classified as the conventional quark model
states ($q\bar{q}$ mesons and $qqq$ baryons) and exotic states
(glueball, hybrid and multiquark states etc.). Hadron spectrum
serves as a golden platform in investigating the low energy strong
interactions. Since the discovery of $X(3872)$ in $2003$ by the
Belle Collaboration~\cite{Choi:2003ue}, many new states in the
charmonium and bottomonium energy regions have been
observed~\cite{Zyla:2020zbs}. Most of these so-called $XYZ$ states
cannot be easily accommodated in the mass spectra of the quark
models, which stimulated the theorists to propose various possible
interpretations of these unconventional
ones~\cite{Chen:2016qju,Guo:2017jvc,Liu:2019zoy,Lebed:2016hpi,Esposito:2016noz,Brambilla:2019esw}.

In the charmonium energy region, two charged charmoniumlike
structures $Z_c(3900)$ and $Z_c(4020)$ were observed by the BESIII
Collaboration in the $J/\psi\pi^\pm$~\cite{Ablikim:2013mio} and
$h_c\pi^\pm$~\cite{Ablikim:2013wzq} channels, respectively. The
$Z_c(3900)$ was subsequently confirmed by the
Belle~\cite{Liu:2013dau} and Xiao {\it et al}~\cite{Xiao:2013iha}.
Latter, the BESIII studied the $(D\bar{D}^\ast)^\pm$ and
$(D^\ast\bar{D}^{\ast})^{\pm,0}$ distributions and found the signals
of $Z_c(3900)$ and $Z_c(4020)$ in the open charmed
channels~\cite{Ablikim:2013xfr,Ablikim:2015swa,Ablikim:2013emm,Ablikim:2015vvn},
respectively. The former was named as the $Z_c(3885)$ because the
mass measured in the $(D\bar{D}^\ast)^\pm$ channel is about $15$ MeV
smaller than that of the $J/\psi\pi^\pm$ channel. Enlightened by the
Ockham's razor: ``{\it Entities should not be multiplied
unnecessarily}", we treat the $Z_c(3900)$ and $Z_c(3885)$ as the
same state that was visualized in different `microscope'. After all,
the mass resolution in different measurements is inequable. In the
bottomonium energy region, the Belle Collaboration discovered two
charged bottomoniumlike states $Z_b(10610)$ and $Z_b(10650)$ in the
$\Upsilon(nS)\pi^\pm$ $(n=1,2,3)$ and $h_b(mP)\pi^\pm$ $(m=1,2)$
invariant mass spectra~\cite{Belle:2011aa}. Four years later, the
Belle Collaboration also observed these two structures in the
$B\bar{B}^\ast$ and $B^\ast\bar{B}^\ast$ channels,
respectively~\cite{Garmash:2015rfd}.

Isospin and parity analyses indicate these $Z_{Q}^{(\prime)}$
$(Q=c,b)$ states are the isovector particles with positive
$G$-parity and negative $C$-parity ($C$-parity for the neutral
members). We will denote the $Z_c(3900),Z_c(4020)$ and
$Z_b(10610),Z_b(10650)$ as $Z_c,Z_c^\prime$ and $Z_b,Z_b^\prime$
respectively in the following context for simplicity. Analyses of
the angular distributions favor the $J^P=1^+$ assignment for the
$Z_c$~\cite{Ablikim:2013xfr,Ablikim:2015swa} and
$Z_b^{(\prime)}$~\cite{Collaboration:2011gja}. The $J^P$ quantum
numbers of the $Z_c^\prime$ are undetermined yet, but the $J^P=1^+$
is presumed in most
works~\cite{Chen:2016qju,Guo:2017jvc,Liu:2019zoy,Lebed:2016hpi,Esposito:2016noz,Brambilla:2019esw}.
The minimal quark component in these $Z_{Q}^{(\prime)}$ states
should be $Q\bar{Q}q\bar{q}$ $(q=u,d)$ rather than the pure
$Q\bar{Q}$ since they are the charged particles. Such a quark
configuration is obviously beyond the conventional mesons and
baryons, so they are dubbed the exotic hadrons. Many theoretical
explanations have been proposed to understand these exotica, such as
the loosely bound molecular states, compact tetraquarks, kinematical
effects and so on (one can consult some comprehensive
reviews~\cite{Chen:2016qju,Guo:2017jvc,Liu:2019zoy,Lebed:2016hpi,Esposito:2016noz,Brambilla:2019esw}
for deepgoing excavations). Besides the similarities of the decay
modes, the mass differences of $(Z_c,Z_c^\prime)$ and
$(Z_b,Z_b^\prime)$ almost equal to the mass splittings of
$(D,D^\ast)$ and $(B,B^\ast)$, respectively. The large comparability
between the $Z_c^{(\prime)}$ and $Z_b^{(\prime)}$ suggests that they
are the partners under the heavy quark flavor symmetry. The most
salient feature of the $Z_c,Z_c^\prime$ and $Z_b,Z_b^\prime$ is
their proximities to the $D\bar{D}^\ast,D^\ast\bar{D}^\ast$ and
$B\bar{B}^\ast,B^\ast\bar{B}^\ast$ thresholds, respectively.
Therefore, the properties of the $Z_{Q}^{(\prime)}$ states are
strongly related to the interactions of these open heavy flavor
systems.

The $Z_c^{(\prime)}$ and $Z_b^{(\prime)}$ lie few MeV above the
$D^\ast\bar{D}^{(\ast)}$ and $B^\ast\bar{B}^{(\ast)}$ thresholds,
respectively. Thus it is natural to investigate whether the
$Z_c^{(\prime)}$ and $Z_b^{(\prime)}$ are molecular resonances
generated from the $D^\ast\bar{D}^{(\ast)}$ and
$B^\ast\bar{B}^{(\ast)}$ interactions, respectively. In this work we
exploit the chiral effective field theory ($\chi$EFT) to study the
$D^\ast\bar{D}^{(\ast)}$ and $B^\ast\bar{B}^{(\ast)}$ interactions
up to the next-to-leading order (NLO), and then fit the experimental
data to extract the resonance parameters. As the modern theory of
nuclear forces~\cite{Weinberg:1990rz,Weinberg:1991um}, $\chi$EFT has
been extensively used to study the nucleon systems with high
precision~\cite{Bernard:1995dp,Epelbaum:2008ga,Machleidt:2011zz,Meissner:2015wva,Hammer:2019poc,RodriguezEntem:2020jgp}.
Within $\chi$EFT, the effective potentials of the $\mathtt{VP}$ and
$\mathtt{VV}$ systems [$\mathtt{V}$ and $\mathtt{P}$ denote the
(anti-)charmed/bottom vector and pseudoscalar mesons, respectively]
with the definite isospin can be respectively parameterized as
\begin{eqnarray}
\mathcal{V}&=&\sum_{i=1}^6 V_i(\bm{p}^\prime,\bm{p})\mathcal{O}_i(\bm{p}^\prime,\bm{p},\bm{\varepsilon},\bm{\varepsilon}^\dagger),\label{VOperator1}\\
\mathcal{V}^\prime&=&\sum_{i=1}^n
V_i^\prime(\bm{p}^\prime,\bm{p})\mathcal{O}_i^\prime(\bm{p}^\prime,\bm{p},\bm{\varepsilon},\bm{\varepsilon}^\dagger,\bm{\varepsilon}^\prime,\bm{\varepsilon}^{\prime\dagger}),\label{VOperator2}
\end{eqnarray}
where $\bm p$ and $\bm p^\prime$ denote the initial and final state
momenta in the center of mass system (c.m.s), respectively.
$\bm{\varepsilon}^{(\prime)}$ and
$\bm{\varepsilon}^{(\prime)\dagger}$ represent the polarization
vectors of the initial and final vector mesons, respectively.
$V_i^{(\prime)}$ are the scalar functions that can be extracted from
the chiral Lagrangians, while $\mathcal{O}_i$ are six pertinent
operators:
\begin{align}
\mathcal{O}_1&=\bm{\varepsilon}^\dagger\cdot\bm{\varepsilon},&\quad\mathcal{O}_2&=(\bm{\varepsilon}^\dagger\times\bm{\varepsilon})(\bm q\times\bm k),\nonumber\\
\mathcal{O}_3&=(\bm q\cdot\bm{\varepsilon}^\dagger)(\bm q\cdot\bm{\varepsilon}),&\quad\mathcal{O}_4&=(\bm k\cdot\bm{\varepsilon}^\dagger)(\bm k\cdot\bm{\varepsilon}),\nonumber\\
\mathcal{O}_5&=(\bm q\times\bm{\varepsilon}^\dagger)(\bm
q\times\bm{\varepsilon}),&\quad\mathcal{O}_6&=(\bm
k\times\bm{\varepsilon}^\dagger)(\bm k\times\bm{\varepsilon}),
\end{align}
with $\bm q=\bm p^\prime-\bm p$ the transferred momentum and $\bm
k=(\bm p^\prime+\bm p)/2$ the average momentum. For the
$\mathtt{VV}$ system, the number of the possible operators increases
drastically due to the involvement of two new polarization vectors
$\bm{\varepsilon}^\prime$ and $\bm{\varepsilon}^{\prime\dagger}$,
e.g.,
\begin{eqnarray}
\mathcal{O}_1^\prime&=&(\bm{\varepsilon}^\dagger\cdot\bm{\varepsilon})(\bm{\varepsilon}^{\prime\dagger}\cdot\bm{\varepsilon}^{\prime}),\qquad~~~\mathcal{O}_2^\prime=(\bm{\varepsilon}^{\prime\dagger}\cdot\bm{\varepsilon})(\bm{\varepsilon}^{\dagger}\cdot\bm{\varepsilon}^{\prime}),\nonumber\\
\mathcal{O}_3^\prime&=&(\bm{\varepsilon}^{\prime\dagger}\cdot\bm{\varepsilon}^{\dagger})(\bm{\varepsilon}\cdot\bm{\varepsilon}^{\prime}),\qquad~~~\mathcal{O}_4^\prime=(\bm{q}\cdot\bm{\varepsilon}^{\prime\dagger})(\bm{q}\cdot\bm{\varepsilon})(\bm{\varepsilon}^{\dagger}\cdot\bm{\varepsilon}^{\prime}),\nonumber\\
\mathcal{O}_5^\prime&=&(\bm{q}\cdot\bm{\varepsilon}^{\dagger})(\bm{q}\cdot\bm{\varepsilon}^\prime)(\bm{\varepsilon}^{\prime\dagger}\cdot\bm{\varepsilon}),\mathcal{O}_6^\prime=(\bm{q}\cdot\bm{\varepsilon}^{\prime\dagger})(\bm{q}\cdot\bm{\varepsilon}^\dagger)(\bm{\varepsilon}^{\prime}\cdot\bm{\varepsilon}),\nonumber\\
\mathcal{O}_7^\prime&=&(\bm{q}\cdot\bm{\varepsilon}^{\prime})(\bm{q}\cdot\bm{\varepsilon})(\bm{\varepsilon}^{\prime\dagger}\cdot\bm{\varepsilon}^{\dagger}),\dots,
\end{eqnarray}
where the ellipsis denotes the other possible combinations among
$\bm q$, $\bm k$, $\bm{\varepsilon}^{(\prime)}$ and
$\bm{\varepsilon}^{(\prime)\dagger}$ at the NLO.

Like the nuclear forces~\cite{Epelbaum:2008ga,Machleidt:2011zz}, the
interactions between a pair of charmed (bottom) mesons can also be
divided into the short-, mid- and long-range contributions. The
$\chi$EFT does not depend on the details of the short-range dynamics
($r\ll 1/m_\pi$), which is usually mimicked by the contact
interaction. Following the spirit of Eq.~\eqref{VOperator1}, the
contact potential of the $\mathtt{VP}$ system is parameterized as
follows,
\begin{eqnarray}\label{Vct}
\mathcal{V}_{\text{ct}}&=&(C_0+C_1\bm q^2+C_2\bm
k^2)\mathcal{O}_1+\sum_{i=2}^{6}C_{i+1}\mathcal{O}_i,
\end{eqnarray}
where $C_i(i=0,\dots,7)$ are the unknown low energy constants
(LECs). The $C_0$ and $C_{1,\dots,7}$ terms designate the leading
order (LO) and the next-to-leading order (NLO) contributions,
respectively. With Eq.~\eqref{VOperator2}, one can construct the
similar form as in Eq.~\eqref{Vct} for the contact potential of the
$\mathtt{VV}$ system.

\begin{figure*}[hptb]
\begin{centering}
    \scalebox{1.0}{\includegraphics[width=1.0\linewidth]{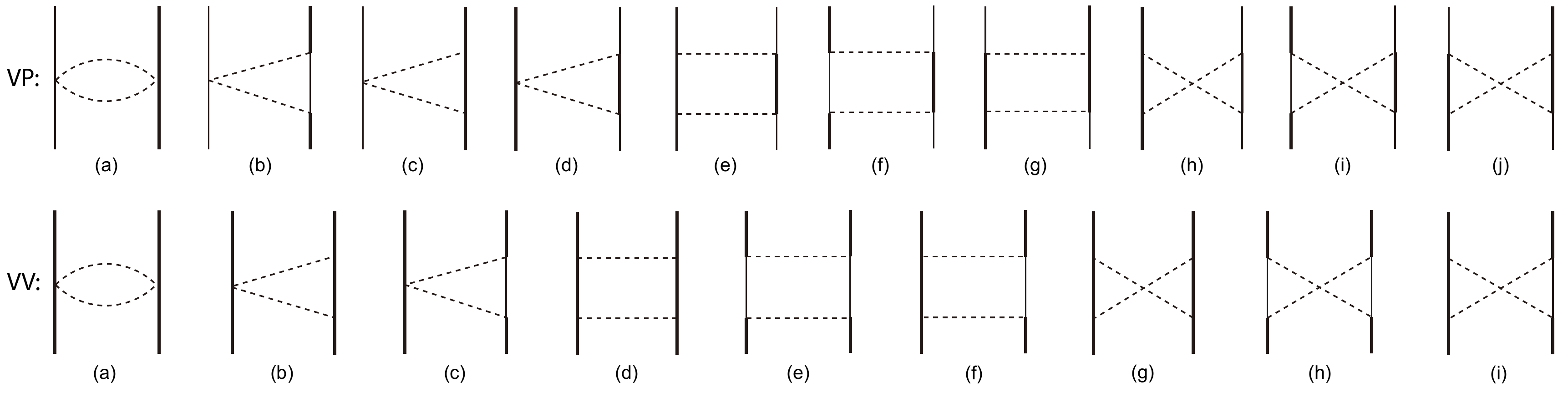}}
    \caption{The two-pion-exchange contributions to the $\mathtt{VP}$ and $\mathtt{VV}$ interactions, where we use the thick, thin and dashed lines to denote the charmed (bottom) vector, pesudoscalar mesons and pion, respectively.\label{TwoPion_Loop1}}
\end{centering}
\end{figure*}

The $\chi$EFT is very good at dealing with the long- and mid-range
interactions, which could be calculated to any high orders
theoretically. For the $\mathtt{VP}$ and $\mathtt{VV}$ systems, the
long-range interaction is provided by the one-pion-exchange (OPE),
which is firmly rooted in the chiral symmetry and its spontaneous
breaking of quantum chromodynamics (QCD). The mid-range force arises
from the two-pion-exchange (TPE). The corresponding loop diagrams
are illustrated in Fig.~\ref{TwoPion_Loop1}. The long- and mid-range
effective potentials can be obtained from the LO chiral Lagrangians,
\begin{eqnarray}\label{LagLO}
\mathcal{L}&=&i\langle\mathcal{H}v\cdot\mathcal{D}\bar{\mathcal{H}}\rangle+g\langle\mathcal{H}\gamma^\mu\gamma_5u_\mu\bar{\mathcal{H}}\rangle\nonumber\\
&&-i\langle
\bar{\tilde{\mathcal{H}}}v\cdot\mathcal{D}\tilde{\mathcal{H}}\rangle+g\langle
\bar{\tilde{\mathcal{H}}}\gamma^\mu\gamma_5u_\mu
\tilde{\mathcal{H}}\rangle,
\end{eqnarray}
where $\langle\cdots\rangle$ denotes the trace in spinor space. The
covariant derivative $\mathcal{D}_\mu=\partial_\mu+\Gamma_\mu$ and
$v=(1,\bm0)$ represents the four-velocity of heavy mesons. The
$\mathcal{H}$ and $\tilde{\mathcal{H}}$ denote the superfield of the
charmed (bottom) mesons and anti-charmed (bottom) mesons,
respectively. Their expressions can be found in
Refs.~\cite{Wise:1992hn,Manohar:2000dt,Wang:2020dhf,Wang:2019nvm}.
The axial coupling $g\simeq0.57$ for the charmed mesons is extracted
from the partial decay width of $D^{\ast+}\to
D^0\pi^+$~\cite{Zyla:2020zbs}, while for the bottom ones average
value $g\simeq0.52$ is taken from the lattice QCD
calculations~\cite{Ohki:2008py,Detmold:2012ge}. The chiral
connection $\Gamma_\mu$ and axial-vector current $u_\mu$ are
formulated as: $\Gamma_\mu\equiv[\xi^\dag,\partial_\mu\xi]/2$, and
$u_\mu\equiv i\{\xi^\dag,\partial_\mu\xi\}/2$, where
$\xi^2=U=\exp\left(i\varphi/f_\pi\right)$, with $\varphi$ the matrix
form of the pion triplet~\cite{Wang:2020dhf}, and $f_\pi=92.4$ MeV
the pion decay constant.

Establishing the flavor wave functions of the
$I^G(J^{PC})=1^+(1^{+-})$ $Z_Q^{(\prime)}$~\cite{Liu:2007bf} and
unfolding Eq.~\eqref{LagLO} one can get the OPE potentials for the
$Z_Q$ and $Z_Q^\prime$ states, respectively,
\begin{eqnarray}
\mathcal{V}_{\text{OPE}}&=&-\frac{g^2}{4f_\pi^2}\frac{\mathcal{O}_3}{\bm q^2+m_\pi^2},\label{VOPE1}\\
\mathcal{V}_{\text{OPE}}^\prime&=&-\frac{g^2}{4f_\pi^2}\frac{(\mathcal{O}_3^\prime-\mathcal{O}_2^\prime)\bm
q^2+\mathcal{O}_4^\prime+\mathcal{O}_5^\prime-\mathcal{O}_6^\prime-\mathcal{O}_7^\prime}{\bm
q^2+m_\pi^2},\label{VOPE2}
\end{eqnarray}
with $m_\pi$ the pion mass, and $\bm
q^2=p^2+p^{\prime2}-2pp^\prime\cos\vartheta$ (where $p=|\bm p|$,
$p^\prime=|\bm p^\prime|$, and $\vartheta$ is the scattering angle
in the c.m.s of $\mathtt{VP}$ and $\mathtt{VV}$). In the Breit
approximation~\cite{Berestetsky:1982bf}, the effective potential
$\mathcal{V}$ from the scattering amplitude $\mathcal{M}$ reads
$\mathcal{V}=-\mathcal{M}/\sqrt{\Pi_i2m_i\Pi_f2m_f}$ ($m_i$ and
$m_f$ stand for the masses of initial and final states,
respectively.).

Similarly, the mid-range potential provided by the loop diagrams in
Fig.~\ref{TwoPion_Loop1} can be calculated with the one-pion and
two-pion coupling vertices in Eq.~\eqref{LagLO} (for the calculation
details one can consult Refs.~\cite{Wang:2018atz,Wang:2019ato}). In
heavy quark limit, the two-particle-irreducible TPE potential can be
formulated via a concise form,
\begin{eqnarray}
\mathcal{V}_{\text{TPE}}^{(\prime)}&=&V_{1}^{(\prime)}\mathcal{O}_1^{(\prime)},\label{VTPE}
\end{eqnarray}
with
\begin{eqnarray}\label{VTPEform}
V_{1}^\prime&=&V_{1}=-\frac{24(4g^2+1)m_\pi^2+(38g^2+5)\bm q^2}{2304\pi^2f_\pi^4}\nonumber\\
&&+\frac{6(6g^2+1)m_\pi^2+(10g^2+1)\bm q^2}{768\pi^2f_\pi^4}\ln\frac{m_\pi^2}{(4\pi f_\pi)^2}\nonumber\\
&&+\frac{4(4g^2+1)m_\pi^2+(10g^2+1)\bm q^2}{384\pi^2f_\pi^4y}\varpi\arctan\frac{y}{\varpi},
\end{eqnarray}
where $\varpi=\sqrt{\bm q^2+4m_\pi^2}$, and
$y=\sqrt{2pp^\prime\cos\vartheta-p^2-p^{\prime2}}$.
\begin{figure}[!hptb]
\begin{centering}
    \scalebox{1.0}{\includegraphics[width=0.95\linewidth]{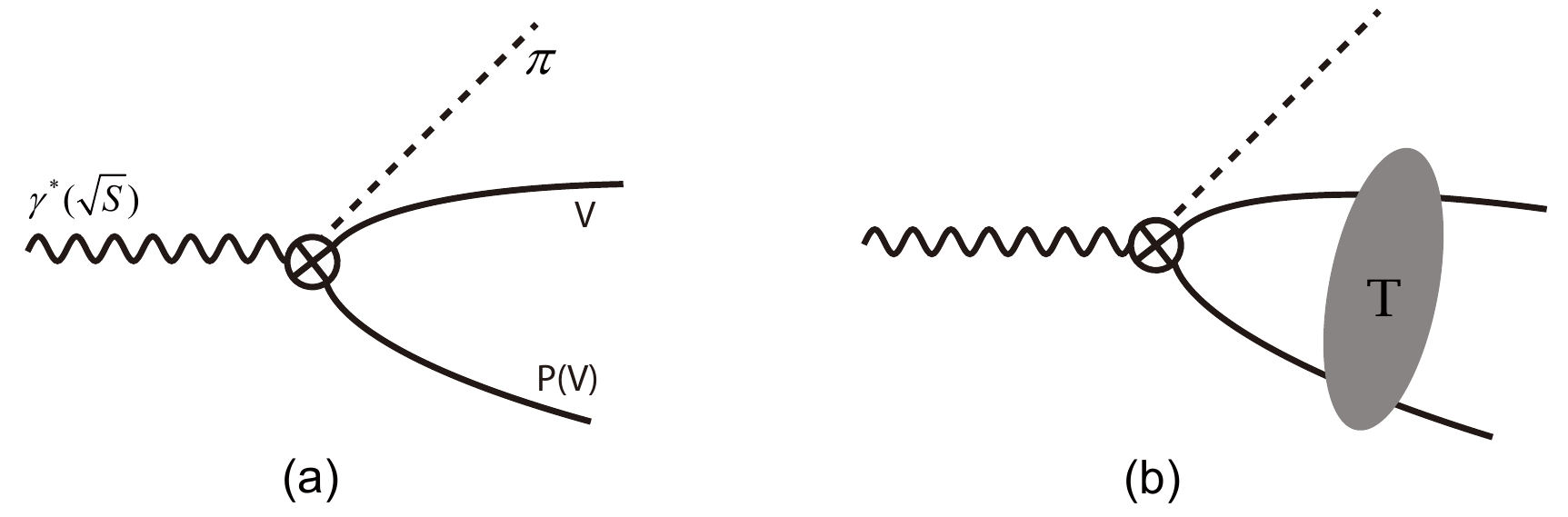}}
    \caption{Graphes (a) and (b) represent the continuum and signal channel contributions, respectively. The wiggly line denotes the virtual photon, and other notations are the same as those in Fig.~\ref{TwoPion_Loop1}. The gray blob in graph (b) signifies the rescatterings of $\mathtt{VP}$ and $\mathtt{VV}$.\label{Production}}
\end{centering}
\end{figure}

The $Z_Q$ and $Z_Q^\prime$ are observed in the $e^+e^-\to\pi
\mathtt{VP}$ and $e^+e^-\to\pi \mathtt{VV}$ processes, respectively.
So we simulate the two transitions and fit the invariant mass
spectra of the $\mathtt{VP}$ and $\mathtt{VV}$ pair. The reaction is
illustrated in Fig.~\ref{Production}, where graphs
\ref{Production}(a) and \ref{Production}(b) describe the continuum
and resonance contributions, respectively. In
Fig.~\ref{Production}(b) we need to cope with the $\mathtt{VP(V)}$
rescatterings, since they account for the dynamical generation of
the $Z_Q^{(\prime)}$. Additionally, we also need to mimic the
$\gamma^\ast\to\pi\mathtt{VP(V)}$ coupling, which can be depicted by
the following effective Lagrangians
\begin{eqnarray}\label{prodvec}
\mathcal{L}_{\gamma^\ast\pi
\mathtt{VP(V)}}&=&g_\gamma\mathcal{F}^{\mu\nu}\mathcal{P}_{\mu\nu}+g_\gamma^\prime\epsilon^{\alpha\beta\mu\nu}\mathcal{F}_{\alpha\beta}\mathcal{P}_{\mu\nu}^\prime
v^\lambda u_\lambda,
\end{eqnarray}
where $g_\gamma^{(\prime)}$ designate the effective coupling
constants, and $\mathcal{F}^{\mu\nu}$ is the field strength tensor
of the virtual photon. $\mathcal{P}_{\mu\nu}^{(\prime)}$ are the
antisymmetric tensors that constructed as:
$\mathcal{P}_{\mu\nu}=(\tilde{P}_\mu^\dagger u_\nu
P^\dagger-\tilde{P}_\nu^\dagger u_\mu P^\dagger)-(\tilde{P}^\dagger
u_\mu P^\dagger_\nu-\tilde{P}^\dagger u_\nu P^\dagger_\mu)$ and
$\mathcal{P}_{\mu\nu}^\prime=\tilde{P}_\mu^\dagger
P_\nu^\dagger-\tilde{P}_\nu^\dagger P_\mu^\dagger$, where $(\tilde{P}_\mu/\tilde{P})$ $P_\mu/P$ denote the
(anti)-charmed (bottom) vector/pseudoscalar meson fields (e.g., see
Refs.~\cite{Wang:2020dhf,Wang:2019nvm}), and $u_\mu$ is the
axial-vector field.

Equipped with the above effective potentials, the $\mathtt{VP}$ and
$\mathtt{VV}$ production amplitudes $\mathcal{U}(E,\bm p)$ can be
obtained by solving the following Lippmann-Schwinger equation (LSE),
\begin{eqnarray}\label{LSE}
\mathcal{U}(E,\bm p)&=&\mathcal{M}(E,\bm p)+\int\frac{d^3\bm q}{(2\pi)^3}\mathcal{V}(E,\bm p,\bm q)\mathcal{G}(E,\bm q)\mathcal{U}(E,\bm q),\nonumber\\
\end{eqnarray}
where $\mathcal{M}(E,\bm p)$ denotes the production vertex from
Eq.~\eqref{prodvec} and $E$ is the invariant mass of the paired
$\mathtt{VP(V)}$. The Green's function $\mathcal{G}(E,\bm q)$ is
given as
\begin{eqnarray}\label{GreenF}
\mathcal{G}(E,\bm q)=\frac{2\mu}{\bm p^2-\bm q^2+i\epsilon},\quad
|\bm p|=\sqrt{2\mu(E-m_{\text{th}})},
\end{eqnarray}
with $\mu$ and $m_{\text{th}}$ the reduced mass and threshold of the
$\mathtt{VP(V)}$ systems, respectively. The potentials in
Eqs.~\eqref{Vct} and \eqref{VOPE1}-\eqref{VTPEform} are given in the
plane wave helicity state basis in the c.m.s of the $\mathtt{VP(V)}$
systems, whereas the physical observables are usually defined in
terms of partial waves, i.e., the $|\ell sj\rangle$ basis (where
$\ell$, $s$ and $j$ represent the orbital angular momentum, total
spin and total angular momentum of the $\mathtt{VP(V)}$ systems,
respectively). So it is desirable to obtain the above effective
potentials in the partial wave decomposition. This can be easily
done via~\cite{Golak:2009ri}
\begin{eqnarray}\label{PWD}
\mathcal{V}_{\ell,\ell^\prime}&=&\int d\hat{\bm p}^\prime\int d\hat{\bm p}\sum_{m_{\ell^\prime}=-\ell^\prime}^{\ell^\prime}\langle \ell^\prime,m_{\ell^\prime};s,m_j-m_{\ell^\prime}|j,m_j\rangle\nonumber\\
&&\times\sum_{m_{\ell}=-\ell}^{\ell}\langle \ell,m_{\ell};s,m_j-m_\ell|j,m_j\rangle\mathcal{Y}_{\ell^\prime m_{\ell^\prime}}^\ast(\theta^\prime,\phi^\prime)\nonumber\\
&&\times\mathcal{Y}_{\ell m_\ell}(\theta,\phi)\langle
s,m_j-m_{\ell^\prime}|\mathcal{V}|s,m_j-m_\ell\rangle,
\end{eqnarray}
with $\mathcal{Y}_{\ell m_\ell}$ the spherical harmonics. The
remaining matrix element $\langle
s,m_j-m_{\ell^\prime}|\mathcal{V}|s,m_j-m_\ell\rangle$ in spin space
can be directly calculated with the coupled spin multiplets
$|1,m_s\rangle$, which are the products of one-body spin states.

As demonstrated in the nucleon systems, the $S$- and $D$-wave mixing
effect plays an important
role~\cite{Bernard:1995dp,Epelbaum:2008ga,Machleidt:2011zz,Meissner:2015wva}.
This effect can be
easily taken into account in the LSE framework, in which the
effective potential becomes a $2\times2$ matrix. After performing
the partial wave decomposition via Eq.~\eqref{PWD}, the contact
potential that incorporates the $S$-$D$ mixing reads,
\begin{eqnarray}\label{VctPWD}
[\mathcal{V}_{\text{ct}}]_{\ell,\ell^\prime}=\left[
\begin{array}{cc}
\tilde{C}_\text{s}+C_\text{s}(p^2+p^{\prime2})&C_\text{sd}p^2\\
C_\text{sd}p^{\prime2}&0
\end{array} \right],
\end{eqnarray}
where $\tilde{C}_\text{s}$, $C_\text{s}$, and $C_\text{sd}$ are the
so-called partial wave LECs. Their values will be fixed by fitting
the experimental data.

Iteration of the potential $\mathcal{V}_{\ell,\ell^\prime}$ in the
LSE requires suppressing the high momenta contribution to avoid
divergence, since the $\chi$EFT is only valid in low momenta region
$q\ll\Lambda_\chi\approx1$ GeV. The Gaussian regulator is commonly
used~\cite{Machleidt:2011zz,RodriguezEntem:2020jgp,Epelbaum:2004fk},
i.e.,
$\mathcal{V}_{\ell,\ell^\prime}\to\mathcal{V}_{\ell,\ell^\prime}\exp(-p^{\prime2}/\Lambda^2-p^{2}/\Lambda^2)$,
where $\Lambda$ is the cutoff parameter. For the nucleon-nucleon
scattering when the high order corrections are
included~\cite{Machleidt:2011zz,RodriguezEntem:2020jgp}, the cutoff
parameter $\Lambda$ is normally chosen to be around $0.5$ GeV. We
leave it as a free parameter and determine its value by fitting the
experimental lineshapes.

In terms of the production amplitude in Eq.~\eqref{LSE}, the
differential decay width for $\gamma^\ast\to\pi \mathtt{VP(V)}$
reads
\begin{eqnarray}\label{diffGamma}
\frac{d\Gamma}{dE}=\frac{1}{12(\sqrt{s})^2(2\pi)^3}|\mathcal{U}(E)|^2|\bm
k_1||\bm k_2^\ast|,
\end{eqnarray}
where $\sqrt{s}$ is the center-of-mass energy of the $e^+e^-$ collision. $\bm k_1$ and $\bm k_2^\ast$ are the three-momentum of the spectator $\pi$ in the c.m.s. of $e^+e^-$ and the three-momentum of $\mathtt{P(V)}$ in the c.m.s of $\mathtt{VP(V)}$, respectively.

\begin{figure*}[!hptb]
\begin{minipage}[t]{0.24\linewidth}
\centering
\includegraphics[width=\columnwidth]{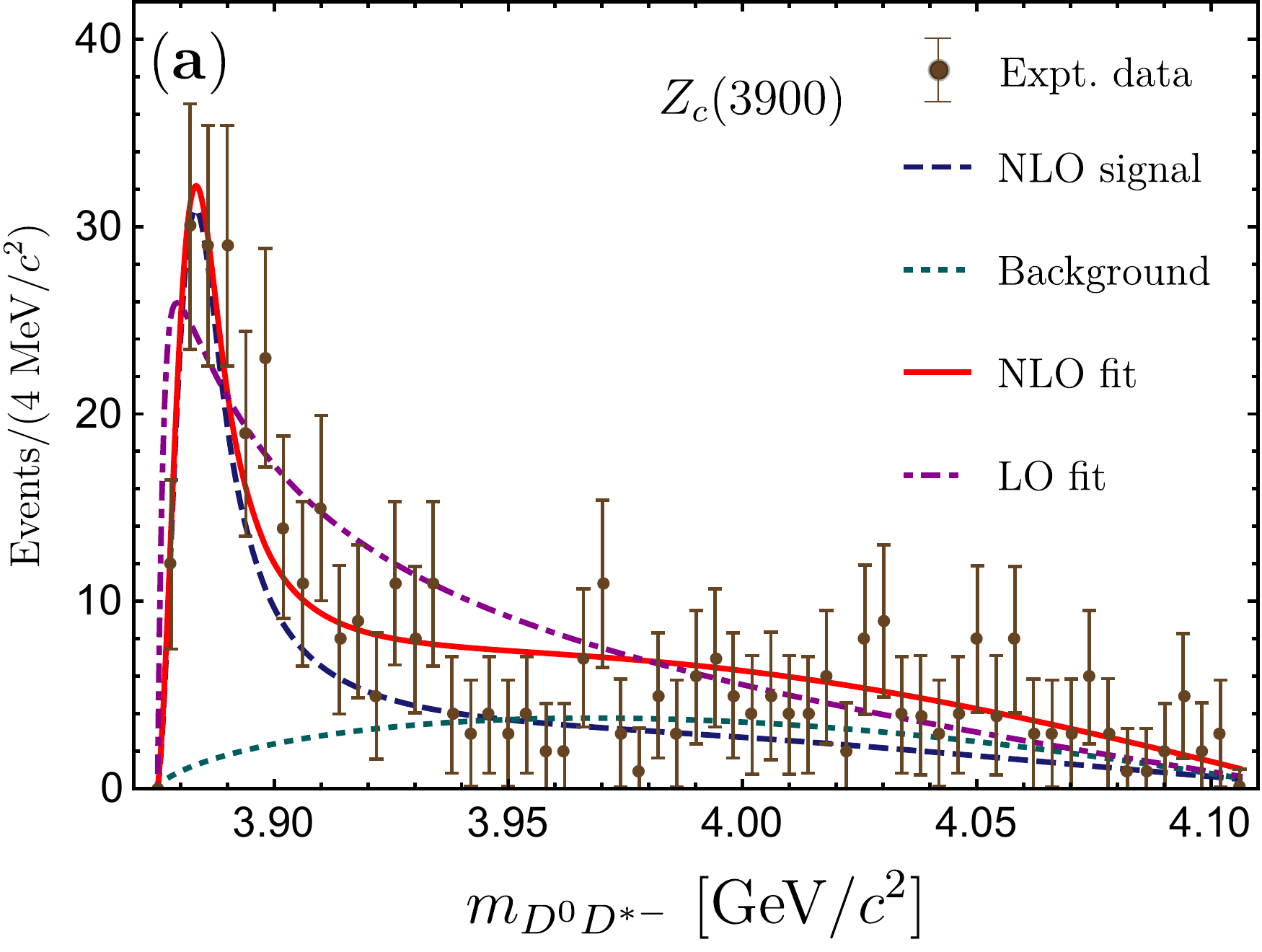}
\end{minipage}%
\hspace{0.05cm}
\begin{minipage}[t]{0.24\linewidth}
\centering
\includegraphics[width=\columnwidth]{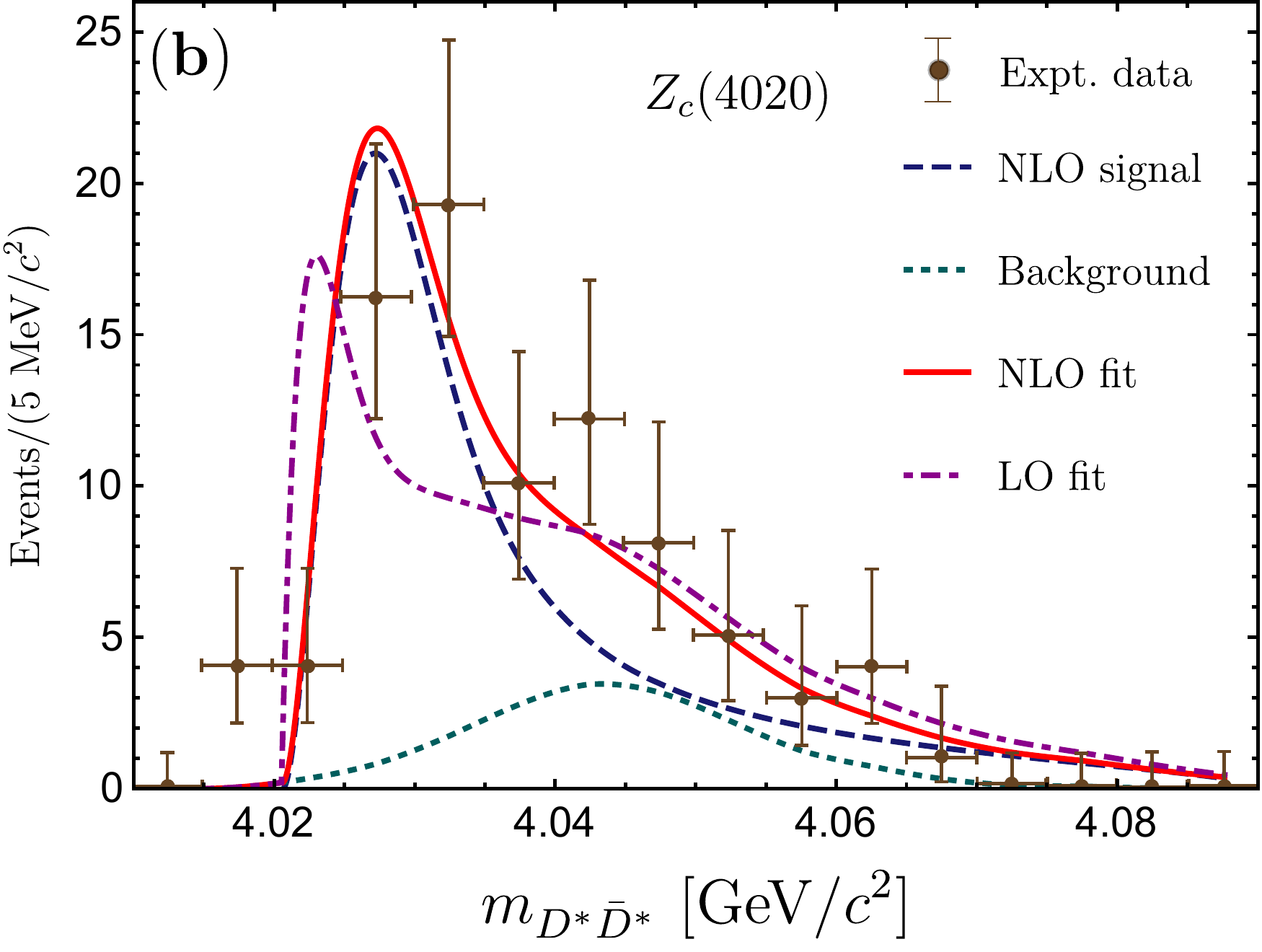}
\end{minipage}
\hspace{0.05cm}
\begin{minipage}[t]{0.252\linewidth}
\centering
\includegraphics[width=\columnwidth]{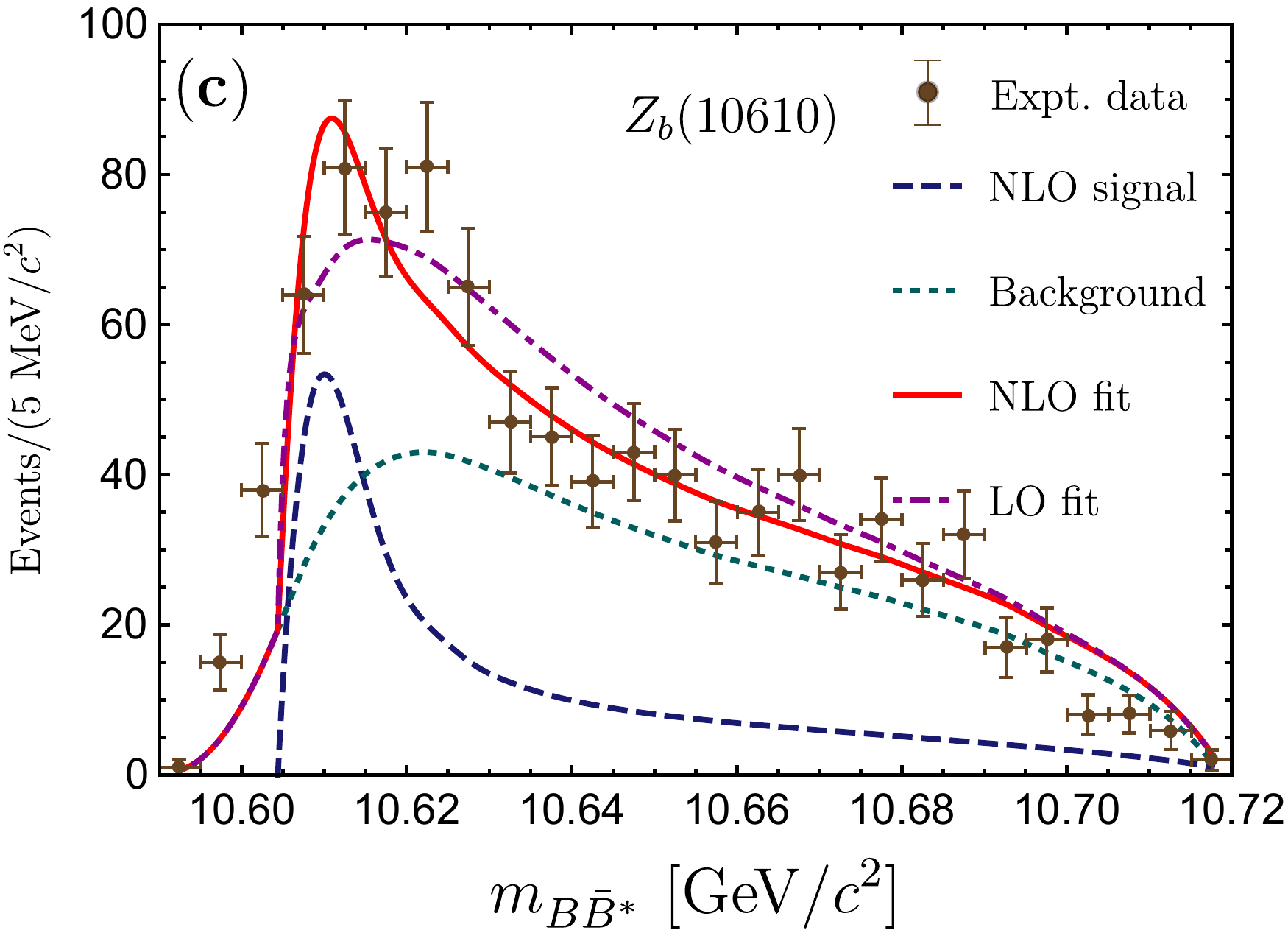}
\end{minipage}
\begin{minipage}[t]{0.246\linewidth}
\centering
\includegraphics[width=\columnwidth]{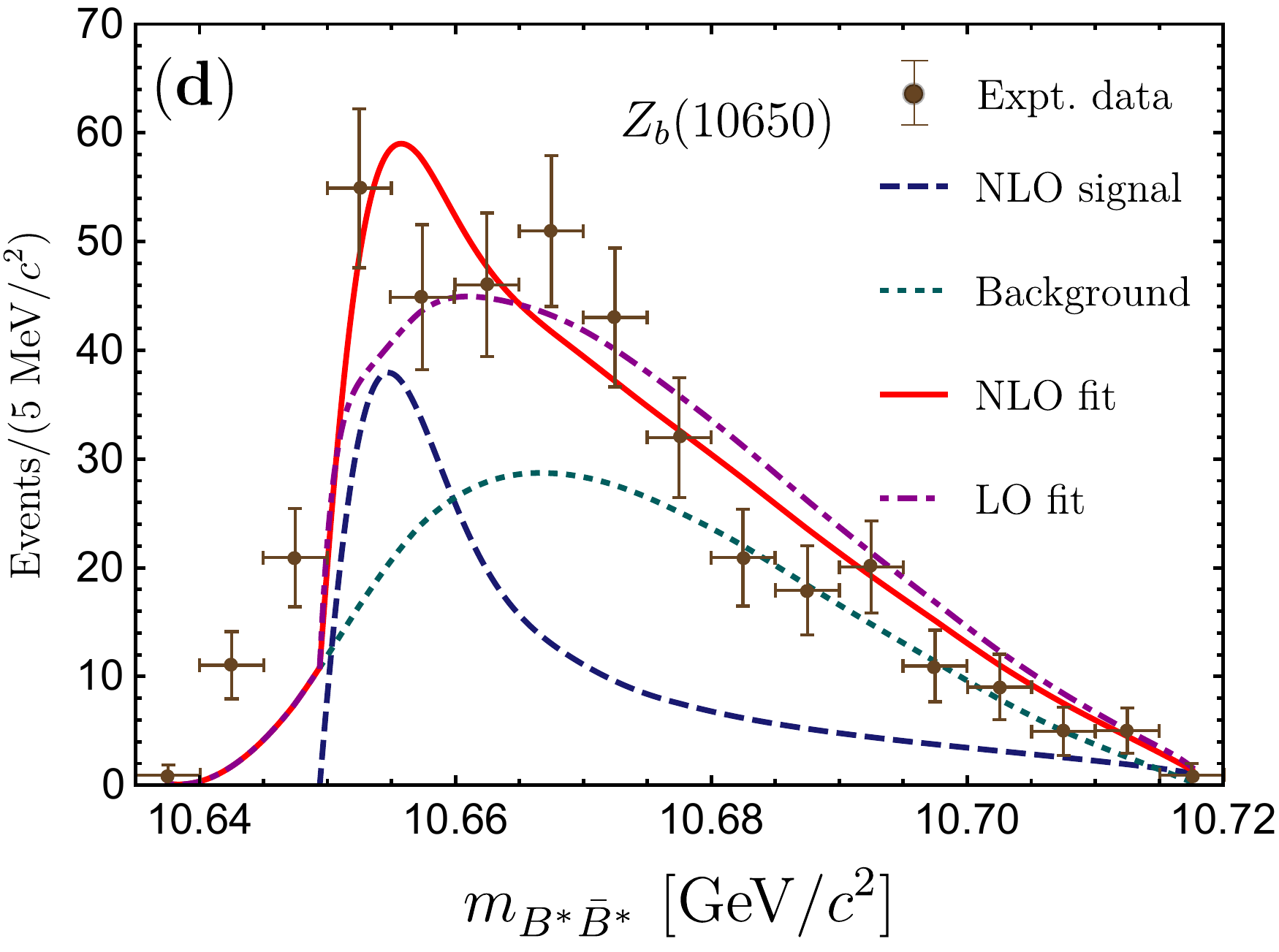}
\end{minipage}
\caption{The $D^\ast \bar{D}^{(\ast)}$ and $B^\ast\bar{B}^{(\ast)}$
invariant mass distributions in $e^+e^-\to\pi\mathtt{VP(V)}$
transitions. The data with error bars in figures (a), (b) and
(c)/(d) are taken from
Refs.~\cite{Ablikim:2015swa},~\cite{Ablikim:2015vvn}
and~\cite{Garmash:2015rfd} at $\sqrt{s}=4.26$, $4.23$, and $10.86$
GeV, respectively. The red solid, dark-blue dashed, purple
dot-dashed and dark-cyan dotted lines denote the NLO fit, NLO
signal, LO fit, and background contributions (extracted from the
corresponding experimental measurements), respectively.
\label{FitZc}}
\end{figure*}

We essentially have four free parameters [three partial wave LECs in
Eq.~\eqref{VctPWD} and a cutoff $\Lambda$] to fit the experimental
lineshapes. For the $Z_c^{(\prime)}$ and $Z_b^{(\prime)}$ states, we
try to fit the $D^\ast \bar{D}^{(\ast)}$ and $B^\ast
\bar{B}^{(\ast)}$ invariant mass distributions measured by the
BESIII~\cite{Ablikim:2015swa,Ablikim:2015vvn} and
Belle~\cite{Garmash:2015rfd} Collaborations, respectively. The
fitted lineshapes and parameters are given in Fig.~\ref{FitZc} and
Table~\ref{FitRes}, respectively. We find the experimental data can
be fitted quantitatively well with the potentials up to the NLO in
our approach. Four sharp peaks appear around $3.88$, $4.02$, $10.61$
and $10.65$ GeV for each distribution, which correspond to the
$Z_c(3900)$, $Z_c(4020)$, $Z_b(10610)$ and $Z_b(10650)$ signals in
experiments, respectively. With the fitted parameters in
Table~\ref{FitRes} as inputs, we search for the poles of the
$T$-matrix in the second (unphysical) Riemann sheet, which can be
achieved through analytical continuation of the Green's function
$\mathcal{G}(p+i\epsilon)$ in Eq.~\eqref{GreenF},
\begin{eqnarray}
\mathcal{G}^{b}(p+i\epsilon)&\equiv&\mathcal{G}^{a}(p+i\epsilon)-2i\text{Im}
\mathcal{G}^{a}(p+i\epsilon),
\end{eqnarray}
where $\mathcal{G}^{a}$ and $\mathcal{G}^{b}$ denote the Green's
function defined in the first (physical) and second Riemann sheet,
respectively.

\begin{table*}[htbp]
\centering
\renewcommand{\arraystretch}{1.5}
\caption{The fitted parameters for the $D^\ast\bar{D}^{(\ast)}$ and
$B^\ast\bar{B}^{(\ast)}$ systems with the potentials up to the NLO,
respectively. The LEC are in units of $10^2$. We define the masses
and widths of the $Z_Q^{(\prime)}$ states from their pole positions
$E=m-i\Gamma/2$ (with $m$ the mass and $\Gamma$ the width). The
masses and widths are given in units of MeV.\label{FitRes}}
\setlength{\tabcolsep}{0.6mm} {
\begin{tabular}{cccccccc}
\hline\hline
States&Thresholds&$\tilde{C}_\text{s}$ [GeV$^{-2}$]&$C_\text{s}$ [GeV$^{-4}$]&$C_\text{sd}$ [GeV$^{-4}$]&$\Lambda$ [GeV]&$[m,\Gamma]_{\text{pole}}$&$[m,\Gamma]_{\text{expt.}}$\\
\hline
$\frac{1}{\sqrt{2}}[D\bar{D}^\ast+D^\ast\bar{D}]$&$3875.8$&$3.6^{+1.2}_{-1.2}$&$-76.9^{+6.2}_{-6.2}$&$1.1^{+5.8}_{-5.8}$&$0.33^{+0.024}_{-0.024}$&$\left[3881.3^{+3.0}_{-3.0},12.4^{+5.0}_{-5.0}\right]$&$\left[3881.7_{-2.3}^{+2.3},26.6_{-3.0}^{+3.0}\right]$~\cite{Ablikim:2015swa}\\
$D^\ast\bar{D}^\ast$&$4017.1$&$4.0^{+1.6}_{-1.6}$&$-78.1^{+8.7}_{-8.7}$&$1.7^{+6.3}_{-6.3}$&$0.34^{+0.031}_{-0.031}$&$[4026.5^{+4.5}_{-4.5},10.1^{+7.2}_{-7.2}]$&$\left[4025.5_{-5.6}^{+3.7},26.0_{-6.0}^{+6.0}\right]$~\cite{Ablikim:2015vvn}\\
$\frac{1}{\sqrt{2}}[B\bar{B}^\ast+B^\ast\bar{B}]$&$10604.4$&$2.2^{+0.2}_{-0.2}$&$-9.9^{+1.0}_{-1.0}$&$3.6^{+4.7}_{-4.7}$&$0.51^{+0.014}_{-0.014}$&$[10607.9^{+2.2}_{-2.2},10.9^{+3.0}_{-3.0}]$&$\left[10607.2^{+2.0}_{-2.0},18.4^{+2.4}_{-2.4}\right]$~\cite{Belle:2011aa}\\
$B^\ast\bar{B}^\ast$&$10649.4$&$2.2^{+0.3}_{-0.3}$&$-9.9^{+1.2}_{-1.2}$&$3.3^{+6.6}_{-6.6}$&$0.51^{+0.015}_{-0.015}$&$[10652.8^{+2.7}_{-2.7},10.9^{+3.4}_{-3.4}]$&$\left[10652.2^{+1.5}_{-1.5},11.5^{+2.2}_{-2.2}\right]$~\cite{Belle:2011aa}\\
\hline\hline
\end{tabular}
}
\end{table*}

We find a pole for each system in the second Riemann sheet with the
pole positions given in Table~\ref{FitRes}. In other words, the
$D^\ast\bar{D}^{(\ast)}$ and $B^\ast\bar{B}^{(\ast)}$ interactions
generate the molecular resonances $Z_c^{(\prime)}$ and
$Z_b^{(\prime)}$. This can be qualitatively understood. When the
$\gamma^\ast$ `emits' a pion, the residual phase spaces for the
$\mathtt{VP(V)}$ systems are small. Thus once the $\mathtt{VP(V)}$
are created near their thresholds, they move slowly and have enough
time to interact with each other. If the interaction is attractive
enough, a bound state is formed, which could not decay into its
component mesons. If the interaction is not attractive enough but
has a barrier to confine the two mesons for a finite time, a
molecular resonance with certain lifetime is produced.

Our extracted masses are all consistent with the experimental
measurements~\cite{Ablikim:2015swa,Ablikim:2015vvn,Garmash:2015rfd},
but the widths in our study are smaller than those of the
experimental data. We do not consider the inelastic channel
$J/\psi\pi~[\Upsilon(nS)\pi]$ and $h_c\pi~[h_b(mP)\pi]$
contributions (see
Refs.~\cite{Hanhart:2015cua,Guo:2016bjq,Wang:2018jlv} for a
couple-channel approach). These inelastic channels would contribute
additional partial decay widths. These inelastic processes occur at
very short distance and cannot be accommodated within the $\chi$EFT
framework. On the other hand, the coupling strength between
$Z_Q^{(\prime)}$ and the inelastic channels is not strong, since the
experimental measurements indicate that the elastic channels
dominate the decay widths of $Z_c$~\cite{Ablikim:2013xfr} and
$Z_b^{(\prime)}$~\cite{Garmash:2015rfd}. Therefore, the corrections
from the inelastic channels to the widths of $Z_Q^{(\prime)}$ shall
not be significant. From Fig.~\ref{FitZc}, the signal lineshapes
deviate from the moderate Breit-Wigner distribution, which are
dramatically distorted by the strong coupling of $\mathtt{VP(V)}$.
The classical Breit-Wigner function is not good enough to describe
these typical very-near-threshold states.

Inspecting the fitted parameters in Table~\ref{FitRes}, one notices
that the rescatterings inside the $\mathtt{VP}$ and $\mathtt{VV}$
systems proceed predominantly via the $S$-wave interactions. They
can be described almost by one set of parameters respectively, which
is guaranteed by the heavy quark spin
symmetry~\cite{Manohar:2000dt,Nieves:2012tt}. In addition, the LO
LEC $\tilde{C}_\text{s}$ for the charmed and bottom systems are
consistent with each other within uncertainties, which is the
reflection of heavy quark flavor
symmetry~\cite{Manohar:2000dt,Bondar:2011ev,Mehen:2011yh}. The
sensible difference of the NLO LEC $C_\text{s}$ for the
$D^\ast\bar{D}^{(\ast)}$ and $B^\ast\bar{B}^{(\ast)}$ systems
encodes the heavy quark flavor symmetry breaking effect. The value
of the cutoff $\Lambda$ also resides in the region
($\Lambda\ll\Lambda_\chi$) where the $\chi$EFT works healthily. The
cutoff for the $B^\ast\bar{B}^{(\ast)}$ systems is larger than that
of the $D^\ast\bar{D}^{(\ast)}$, since the interaction radius
($R\sim1/\Lambda$) for the $B^\ast\bar{B}^{(\ast)}$ is shorter than
that of the $D^\ast\bar{D}^{(\ast)}$. It is well known that the
bottom mesons are heavier than the charmed ones.

We also attempt to fit the data with the LO effective potentials
solely (OPE plus the LO contact terms), but cannot reproduce the
experimental lineshapes well (purple dot-dashed lines in
Fig.~\ref{FitZc}). Those bumps are caused by the sudden opening of
the phase spaces together with the monotone decreasing behavior of
the production amplitudes, but not by any genuine poles of the
$T$-matrix in the second Riemann sheet. These signals become bound
states with the LO interaction. Nevertheless, the parameters
obtained with only the LO interaction are less reasonable, such as
$\tilde{C}_\text{s}\simeq-134.8$ GeV$^{-2}$ and $\Lambda\simeq1.37$
GeV for the $Z_c^{(\prime)}$ states (while
$\tilde{C}_\text{s}\simeq-29.3$ GeV$^{-2}$ and $\Lambda\simeq1.43$
GeV for the $Z_b^{(\prime)}$ states). Although there are no
guidances to judge the values of $\tilde{C}_\text{s}$, the $\chi$EFT
imposes strong constrains to the $\Lambda$, which has to be smaller
than the typical hard scale, i.e., the $\rho$ meson mass
$m_\rho\simeq0.77$ GeV. Therefore, we can conclude that either from
the fitting quality or the rationality of parameters, the bound
state explanations are not favored.

As elucidated above, the $Z_Q^{(\prime)}$ states can be well
identified as the molecular resonances. In the resonance scenario,
their decay behaviors can be explained qualitatively well. In
contrast to the bound state, a resonance naturally dissolves to
their components after interacting within finite time, which
contributes to the dominant decay mode. The decays with final states
of a heavy quarkonium and a light meson,
$[Q\bar{q}]+[\bar{Q}q]\to[Q\bar{Q}]+[q\bar{q}]$ proceeds with less
probability, which are induced by much shorter range interaction
(compared to $1/\Lambda_{\chi}$). At the hadron level, these decays
take place via exchanging a heavy meson $[Q\bar{q}]$, which is
generally suppressed. This is why the partial widths from the
inelastic channel contributions are much smaller than those of the
elastic channels in
experiments~\cite{Ablikim:2013xfr,Garmash:2015rfd}.

In summary, we systematically study the $D^\ast\bar{D}^{(\ast)}$ and
$B^\ast\bar{B}^{(\ast)}$ effective potentials with the $\chi$EFT up
to the NLO to draw a clear picture of their interactions. With these
potentials, we investigate the internal structures of the
experimentally observed $Z_c^{(\prime)}$ and $Z_b^{(\prime)}$ states
in recent years. The short-, mid- and long-range forces are all
included to fit the invariant mass distributions. The experimental
data are fitted very well with the effective potentials up to the
NLO. The peaks in experiments arise from the poles in the second
Riemann sheet, which indicate the $Z_c^{(\prime)}$ and
$Z_b^{(\prime)}$ states are resonances that are generated from the
analogue of nuclear forces in heavy meson sectors. The heavy quark
symmetry and its breaking effect are both reflected in the
parameters. The fittings with the LO potentials give rise to the
bound states, which is repudiated either by the above-threshold
masses or the validity of $\chi$EFT. The decay behaviors of the
$Z_c^{(\prime)}$ and $Z_b^{(\prime)}$ states can also be
qualitatively interpreted in the resonance picture. In our study,
the $Z_Q^{(\prime)}$ signals can be fully reproduced by the
$\mathtt{VP(V)}$ rescatterings, where the initial states
$\pi\mathtt{VP(V)}$ are assumed to be produced from point-like
sources. We do not need additional structures around the colliding
energies.

Besides the $XYZ$ states, more and more new states have been
observed in experiments (such as the $P_c$~\cite{Aaij:2019vzc} and
very recently reported $X_{0,1}$ states at
LHCb~\cite{Johnson:2020xx}), thus a model independent way is
urgently called for to illuminate the nature of these new hadrons.
The systematical generalization of the $\chi$EFT to the heavy meson
systems is very successful in this work, which helps us to pin down
the inner structures of the $Z_c^{(\prime)}$ and $Z_b^{(\prime)}$
states. This framework can also be applied to investigate whether
the other near-threshold states (e.g., $P_c$ and $X_{0,1}$) have the 
same origin, i.e., the dynamically generated resonances (bound states) 
from the analogue of nuclear forces in different sectors. 
This would undoubtedly deepen our understandings of the low energy 
behaviors of QCD.

\medskip
This project is supported by the National Natural Science Foundation
of China under Grant 11975033.

\end{document}